# Stain-free Detection of Embryo Polarization using Deep Learning

Cheng Shen[1,+], Adiyant Lamba[2,+], Meng Zhu[2,3], Ray Zhang[4], Changhuei Yang[1,5,*], and Magdalena Zernicka Goetz[2,5,*]

[1] Department of Electrical Engineering, California Institute of Technology, Pasadena, CA, USA

[2] Mammalian Embryo and Stem Cell Group, Department of Physiology, Development and Neuroscience, University of Cambridge, Downing Street, Cambridge, CB2 3EG, UK

[3] Blavatnik Institute, Harvard Medical School, Department of Genetics, Boston, MA 02115, USA

[4] Department of Pathology and Immunology, Washington University School of Medicine, St. Louis, MO, USA

[5] Division of Biology and Biological Engineering, California Institute of Technology, Pasadena, CA, USA

[+] These authors contributed equally

[*] co-corresponding authors: chyang@caltech.edu, magdaz@caltech.edu

## Abstract

Polarization of the mammalian embryo at the right developmental time is critical for its development to term and would be valuable in assessing the potential of human embryos. However, tracking polarization requires invasive fluorescence staining, impermissible in the in vitro fertilization clinic. Here, we report the use of artificial intelligence to detect polarization from unstained time-lapse movies of mouse embryos. We assembled a dataset of bright-field movie frames from 8-cell-stage embryos, side-by-side with corresponding images of fluorescent markers of cell polarization. We then used an ensemble learning model to detect whether any bright-field frame showed an embryo before or after onset of polarization. Our resulting model has an accuracy of 85% for detecting polarization, significantly outperforming human volunteers trained on the same data (61% accuracy). We discovered that our self-learning model focuses upon the angle between cells as one known cue for compaction, which precedes polarization, but it outperforms the use of this cue alone. By compressing three-dimensional time-lapsed image data into two-dimensions, we are able to reduce data to an easily manageable size for deep learning processing. In conclusion, we describe a method for detecting a key developmental feature of embryo development that avoids clinically impermissible fluorescence staining.

## Introduction

Mammalian embryo polarization is the process by which all individual cells of the embryo establish an apical domain on the cell-cell contact-free surface. In the mouse embryo, this process occurs at the late 8-cell stage, on the third day of development after fertilization, (Fig. 1a) and in humans on the fourth day at the 8-16 cell stage[1-7]. The apical domain is composed of the PAR complex and ERM proteins (Ezrin, Radixin, Moesin), enclosed by an actomyosin ring[2, 8-12]. The cells which inherit this apical domain after division will become specified as trophectoderm (TE), which ultimately forms the placenta. In contrast, those cells that do not inherit the apical domain will form the inner cell mass (ICM), which will give rise to all fetal tissues and the yolk sac[4-6]. Thus, embryo polarization provides the first critical bifurcation of cell fates in the mammalian embryo, and establishment of cell lineages in the blastocyst, which is crucial for implantation and a successful pregnancy. In agreement with this, preventing cell polarization of the mouse and human embryo, prevents its successful development[3, 4, 11, 13].

Given the importance of polarization, an ability to detect this developmental feature non-invasively would be beneficial, for example, for the screening of viable human embryos for implantation. However, all current methods for detecting polarization are invasive as they rely on modifying embryos to express fluorescently tagged proteins that mark the apical domains[14, 15]. Such fluorescent tagging of human embryos meant for implantation is impermissible, which currently prevents clinical embryologists from utilizing polarization to evaluate the quality of human embryos for transfer to mothers-to-be.

Tracking polarization without the use of fluorescence could be solved using deep learning, which is able to discern salient features that may be unintuitive for humans[16, 17]. Indeed, deep learning has been recently used successfully to automate detection of an embryo's morphological features and applied on single time-point images to assess implantation potential of human embryos[16-21]. These deep learning approaches either provide a means of accurately counting cell numbers with developmental time[16, 17]; relate embryo morphology to subjective developmental criteria assessed by embryologists[18, 19]; or provide independent assessment of morphological features[19, 20]. One study has related preimplantation morphology with successful development of fetal heartbeat[21]. However, the morphological features being assessed by the deep learning algorithms used to date are generally not clear. In addition, these current approaches do not rely upon known critical developmental milestones in the molecular cell biology of preimplantation development. Here, we have used time lapse movies of fluorescent markers of polarization in the developing mouse embryo to train a deep learning system to recognize the polarization events in the corresponding bright field movie frames with a high degree of success. This is the first time that deep learning has been applied to recognize a specific molecular cell biological process in an embryo that is key for developmental success.

## Results

### Collection and annotation of embryo images

In order to develop our deep learning model for detecting the polarization status of live and unstained embryos, we first required a large dataset of DIC embryo frames for which the polarization is unambiguously evident. Generating this dataset required each DIC image to have a corresponding channel that uses fluorescently tagged proteins to indicate polarization for each embryo clearly. The polarization of a single blastomere in the embryo can be determined by the localization of apical proteins, which are enclosed by an actomyosin ring[9-11, 22]. We built a large dataset composed of synchronized DIC and fluorescence channels of mouse embryos during polarization by collecting mouse embryo time-lapse recordings, each containing a fluorescence channel to indicate embryo polarization, and a DIC channel for model training and testing (Fig. 1b). For time-lapse recordings, embryos were injected at the 2-cell stage with synthetic mRNA for Ezrin tagged with red fluorescence protein (RFP), as previously[23], and cultured in vitro to the 16-cell stage. We used Ezrin as a marker for blastomere polarization, as Ezrin localizes to the apical surface during the formation of an apical polarity domain[12, 24]. Using the Ezrin-RFP fluorescent channel, we determined the time point at which the first blastomere of the embryo polarized for each time-lapse recording, indicated by formation of a clear apical polarity cap (Fig. 1c, Supplementary Fig. 1). Using this annotation, each DIC frame was labelled as either before or after the onset of polarization (Fig. 1d). In total, we produced a dataset containing 89 embryo time-lapse recordings of the 8-cell stage embryo during their polarization.

**Compression of 3D embryo image sequences**

In previous studies, a single slice image along the z axis was used for model input[16-21] due to the use of existing deep learning models designed for a two-dimensional (2D) image input. However, a single z-slice image does not capture 3D embryo structural information. Analysis of a 3D image stack with deep learning requires a re-designed model architecture that dramatically increases the complexity and time required for model development[25, 26]. Moreover, adapting existing pre-trained deep learning networks for 3D analysis through transfer learning[27] would not be straightforward as these networks are predominantly designed for 2D image recognition tasks. To resolve this problem, we utilized a state-of-the-art all-in-focus (AIF) algorithm based on dual-tree complex wavelet transform (DTCWT)[28] to compress the optically sectioned z stack of each DIC frame in our dataset. The result was a single 2D AIF DIC image capturing the majority of relevant but sparsely distributed 3D embryo information at each time point (Fig. 1b).

We found that AIF images based on DTCWT could reveal all blastomeres of a 3D embryo in a single 2D image (Supplementary Fig. 2). In contrast, the median z slice typically contained several blastomeres that were optically out of focus, resulting in lost information. AIF images also resembled standard images, allowing for straightforward transfer learning using open-source 2D image classification models pre-trained on ImageNet[29] as initialization.

**Model architecture**

The dataset consisting of AIF DIC images paired with corresponding annotated polarization labels was randomly split into a training cohort of 70 embryos (1889 frames) and a testing cohort of 19 embryos (583 frames) (Fig. 2a). These were used as learning and evaluation datasets, respectively, for a single deep convolutional neural network (DCNN) binary classification model. For supervised learning of DCNN models, we retained only information about whether a frame was before or after onset and stripped away other time information (Fig. 1d, 2b). On individual testing frames, each DCNN model outputs whether or not polarization was detected as a vector containing two probabilities—one for each class (before or after onset, Fig. 2b). To mitigate over-fitting, we ensembled six DCNN models trained using different initializations and different optimizers but trained over the same number of epochs. The final polarization status prediction for a single input image is the class (before or after onset) having the highest average probability across all six contributing models. Overall, our model accuracy increased from an average of 82.6% for a single DCNN to 85.2% with ensemble learning.

**Ensemble deep learning model outperforms human volunteers**

We recruited six volunteers following the criteria (outlined in the Methods section), to compare polarization detection accuracy against our model. We aimed to recruit human volunteers from a STEM background, who would be motivated to benefit from the technology in a clinical setting and who might compare favorably with our machine learning system. The volunteers were self-trained using the same annotated training dataset used by our model. They were then given the same AIF DIC testing dataset and asked to determine the polarization status for each test image (before or after onset).

The model we establish here yielded a classification sensitivity of 90.2% (95% confidence interval (CI): 86.1% - 93.8%) and specificity of 81.1% (95% CI: 76.2% - 85.4%) for single image inputs, with areas under the receiver operating characteristic curve of 0.893 (95% CI: 0.866 - 0.917) (Fig. 3a and Supplementary Table 1). Deep learning achieved both a higher true positive rate and lower false positive rate than the average human volunteer. Figure 3b shows the confusion matrix for predictions. Our model correctly classified 497 out of 583 frames, resulting in a classification accuracy of 85.2% (95% CI: 82.2% - 88.2%). In comparison, the average human accuracy on the same testing frames was 61.1% (95% CI: 57.1% - 65.0%) (Fig. 3b). The model outperformed humans on average (Fig. 3c, two-tailed z-test, $p<0.0001$) as well as individually (Supplementary Fig. 3).

**Understanding image features of interest to the model**

We interrogated our model for embryo regions that most strongly affected the model's predictions, using class activation maps (CAM)[30]. CAM relies on a heat map representation to highlight pixels that trigger a model to associate an image with a particular class (before or after onset). In Fig. 4, we have overlaid the CAM heat map with the input testing AIF DIC image. In each heat map, red pixels indicate regions of the embryo containing features that correlate positively with the predicted polarization class, while blue pixels indicate regions containing features that correlate negatively (i.e., correlate positively with the opposing class). To understand which regions of an embryo influence our model most, we evaluated each possible prediction outcome: true negative (TN) (Fig. 4a), false positive (FP) (Fig. 4b), false negative (FN) (Fig. 4c), and true positive (TP) (Fig. 4d). When the model classified image frames as after polarization, it appeared to use inter-blastomere angle as a cue (see Discussion). Misclassifications tended to result from mismatched polarity between individual blastomeres and the overall embryo, producing weak prediction probabilities for both classes near 50% while the model was forced to choose one class (Fig. 4c). Predictions in this probability range are more reasonably interpreted as not sure or cannot tell, but these were not options for the model.

**Model outperforms compaction alone for discrimination**

The use of inter-blastomere angle as a cue by our model to determine embryo polarization (Fig. 4a) was not surprising. Inter-blastomere angle is an indicator of embryo compaction[31, 32], a morphological change during development that typically precedes polarization (Fig. 5a). To assess the extent to which our deep learning model uses just compaction for its polarization prediction, we annotated each embryo's AIF DIC frame sequence with the time point of compaction. We defined the time of compaction as the first frame at which the smallest inter-blastomere angle of the embryo is over 120 degrees, in agreement with previous research (Supplementary Fig. 4)[32]. To find the model's predicted time point of polarization, we re-aligned embryo frames in their original time sequence and applied temporal smoothing on the predicted label sequence for each testing embryo based on majority voting to output a single time point for polarization (Fig. 2c).

The Pearson correlation coefficient between compaction time point and the model's predicted time point of polarization onset was 0.75 across the 19 embryos used for testing (Fig. 5b), suggesting that whilst compaction is indeed a utilized cue, it is not the only factor used by the model. We evaluated whether our model was superior to using compaction alone as a proxy for polarization, by calculating the time discrepancies between annotated polarization time indexes (ground truth) and predicted time indexes by either our model or the compaction proxy. The model had significantly smaller time point prediction errors compared to the latter (two-tailed Wilcoxon matched-pairs signed-rank test, $p<0.05$, Fig. 5c). That is, the model was superior to the use of compaction alone for predicting polarization and has likely managed to learn additional cues we do not yet understand.

**Distinguishing exact polarization onset time**

We wished to further extend our deep learning model to identify the exact point at which polarization occurs in time-sequence videos. To this end, we evaluated polarization onset time point predictions from the classification results of both the model and human volunteers, using a temporal smoothing method (Fig. 2c). Timestamp errors between predicted and annotated time points were calculated as was done previously for compaction time point analysis. Our model had significantly smaller timestamp prediction errors than the average human volunteer by pairwise comparison (two-tailed Wilcoxon matched-pairs signed-rank test, $p<0.01$, Fig. 6; Supplementary Fig. 5a).

We next wished to investigate whether smoothened results from our ensemble classification model could outperform even human volunteers who are given access to temporal information during testing that the model does not use. To this end, we provided each volunteer with the complete AIF DIC videos in frame-by-frame time sequence for each embryo and asked for their estimate of the polarization onset time point. Compared with the smoothened model classification results performed on individual unordered images, the average human timestamp discrepancy was significantly larger than that of our model (two-tailed Wilcoxon signed-rank test, $p<0.05$, Fig. 6; Supplementary Fig. 5b). The model identified exact polarization time points more precisely than the human volunteers, even when the volunteers utilized temporally ordered full video frames that the model did not have access to during training.

**Discussion**

In this study, we show that an ensemble deep learning model can identify polarization in unstained embryo images from the DIC microscope with an accuracy surpassing that of humans by a wide margin. When classifying 583 test DIC 8-cell stage frames, our model yielded an accuracy of 85% [95% confidence interval (CI): 82.2% - 88.2%] compared to corresponding average human accuracy of 61% [95% CI: 57.1% - 65.0%].

It is important to note the difficulty of the polarization detection task using unstained embryo images, since to the naked human eye, unstained images do not have any clear features which allow identification of the cellular apical domain. This is reflected in our observed human accuracy of 61%, which represents a performance level barely higher than random chance. Expressed as odds, the odds of a human volunteer correctly differentiating polarization were 1.5—that is, humans were right 1.5 times for each time they were wrong. In contrast, our deep learning model was right 5.7 times for each time it was wrong.

Current embryo selection in IVF clinics relies on crude and qualitative expert inspection of live embryos under plain microscopy that equates to an educated guess. Deep learning is an unusually well-suited solution to providing a more accurate assessment of embryo health for IVF, since deep neural networks recognize subtle features that are difficult for humans to identify[33-35]. Prior research in this field[16, 17, 20] limited itself only to features that are obvious on bright field or DIC imaging such as cell count and size, or to directly predict implantation potential without investigating underlying biological processes[19]. Our model can potentially enable embryo quality assessment using an important developmental milestone and thereby overcome some limitations of these prior deep learning studies. To our knowledge, there is currently no other known way to adequately evaluate the developmentally critical polarization milestone for embryo health screening prior to selection for implantation. By detecting an underlying developmental feature of the embryo using unstained embryo images, our study provides a platform for a potential future solution to improve IVF technology.

We investigated possible reasons for the successes and failures of our model using the CAM technique and concluded that inter-blastomere angle, an indicator of compaction, was one of the model's cues for prediction. However, compaction alone was an inferior predictor of polarization compared to the model, suggesting that our model learned additional features informative of polarization that we currently do not understand. The intriguing implication is that more discriminative biology is apparent in simple unstained embryo images than we currently realize. Moreover, our deep learning model was able to identify the exact time point of polarization onset amongst temporally sequenced video frames better than all human volunteers, even with a severe disadvantage in data.

We were able to circumvent 3D image stack analysis through the use of a state-of-the-art all-in-focus algorithm[28], which allowed for the efficient collapse of 3D optical data to 2D. Prior studies that apply deep learning to embryo development have used single z slice DIC images as input, which obscures important 3D blastomere features. Our DTCWT-based method of compressing multiple z slices into a single maximally informative 2D representation reduces data size, allowing a full range of published 2D neural network models to become accessible for analysis.

We considered several possibilities during the design of our model architecture. Many DCNN architectures have been published for image classification, including dense convolutional network[36], squeeze-and-excitation network[37] and residual neural network[38]. We used cross validation to select a final candidate architecture among them these networks and found that ResNet-50-v2 has the highest average validation accuracy and a low variance (Supplementary Fig. 6). This model became the basis for our individual DCNN modules, of which we combined six to form the final ensemble polarization classifier. Here the number six is to match the number of recruited human volunteers for fair comparison.

Our deep learning-based approach presents a significantly more accurate and less laborious way to evaluate mammalian embryo polarization compared to manual analysis. In future studies, this approach can be used for stainless tracking of polarization in live human embryos, allowing IVF clinics to assess embryo polarity status and its precise timing using non-invasive methods, and move away from empirical embryo grading to a system grounded in established developmental milestones of embryogenesis.

In conclusion, we have developed a powerful non-invasive deep learning method to detect embryo polarization from images without the use of fluorescence, while surpassing human performance. This method has a great potential to provide the first example of detecting an underlying developmental feature of the mammalian embryo from unstained images, which could be an important step towards improving IVF technology from the rate of increase in cell number or assessment of morphological features independently of developmental events.

## Methods

### Assembling the embryo dataset

All mouse experimental data was obtained in accordance with the Animals (Scientific Procedures) Act 1986 Amendment Regulations 2012, under project license by the University of Cambridge Animal Welfare and Ethical Review Body (AWERB). Reporting of animal experiments follows ARRIVE guidelines. Embryos were collected at the 2-cell stage from F1 females (C57BI6xCBA) mated with F1 studs, following super ovulation of the female: injection of 7.5 IU of pregnant mares' serum gonadotropin (PMSG; Intervet), followed by injection of 7.5 IU of human chorionic gonadotropin (HCG; Intervet) after 48 hours and immediately prior to mating.

Embryos were microinjected with Ezrin–red fluorescent protein (RFP) mRNA as a polarity marker before imaging, in each blastomere at the 2-cell stage, as described previously[23]. Images were collected on confocal Leica SP5 or SP8 microscopes. The interval between each frame on the time ($t$) axis was 1200 s - 2400 s for each embryo, and $z$ frames were taken at 4 $\mu$m intervals on the z axis. Time-lapse recordings were converted into TIFF files for analysis and processed on Fiji software. Recordings that were incorrectly formatted, visually unclear, or which showed grossly defective embryos were excluded. From an initial 174 embryo recordings, 89 were used for deep learning and human testing (Supplementary Fig. 1). Only 8-cell stage frames were included in deep learning and analysis (defined as frames from the first frame where 8 distinct blastomeres are visible, to the frame immediately prior to the moment at which the final blastomere starts dividing). The DIC channel images were converted into an AIF DIC frame for each time point as described in Results, and the Ezrin-RFP channel images were converted into maximum intensity $z$ projection frames, prior to annotation.

### Embryo annotation (polarization and compaction)

Each embryo time-lapse recording was marked with a polarization onset time by a human expert annotator, corresponding to the first frame in which a polarized blastomere is clearly visible. This was achieved using the maximum intensity z projection Ezrin-RFP frame: the polarization onset frame is one in which the first apical ring or cap is completely and clearly formed (closed) on any blastomere, and which takes up greater than or equal to 1/3 of the surface of the cell as visible in the recording. All frames after and including this polarization onset point were classified as after-onset. All frames prior to this point were classified as before-onset. Compaction time was indicated when smallest inter-blastomere angle was greater than 120 degrees, as previously[32]. All frames after and including this point were considered compacted, and all frames prior to this point were considered uncompacted.

**Ensemble deep learning framework**

In this study, two types of effective machine learning techniques, DCNN and ensemble learning, were adopted and combined together for prediction of polarity onset. Multiple (6 here to match the number of human volunteers) DCNNs learnt on the training cohort and then their output predictions were averaged to predict the class label of each testing image. Specifically, the ResNet backbone was chosen as the main part of each DCNN model. A dense layer with two output nodes is added on top of the ResNet backbone. We used the pre-trained weights on ImageNet database as the initialization for each DCNN model. Three of them were trained with SGD optimizer and the other three were trained with Adam optimizer. All of them were trained for 40 epochs. At the end of 40 epochs, all the models converge to nearly 100% in terms of the training accuracy. Different training settings made the six trained CNNs a bit more diverse from each other, where the diversity among CNNs would improve the generalization ability of the ensemble model. We also adopted the cross-validation technique to compare with other backbones, including DenseNet and SENet. Based on the results of 5-fold CV experiments, we found ResNet is the optimal choice in both prediction performance and computational load.

**Human trial**

In order to evaluate the performance of our DL model, comparative trials on human volunteers to identify polarity onset were conducted as well. Six human volunteers (3 males, 3 females for gender equality) with a bachelor’s degree in a STEM subject but without prior experience of mouse embryo development studies were recruited from Caltech community, as representatives for competent STEM-trained but inexperienced volunteers who would benefit from the technology in a clinical setting. Volunteers were sent an email with clear instructions and a link to the training and testing data. Each was asked to learn on the training dataset first and then apply their learnt patterns to the testing images, to predict their polarity onset status by filling in an Excel table with predicted labels. After the test, they each returned their Excel file for evaluation.

All participants provided informed consent before taking part in our study. They consented to allow their data to be used in the final analysis and all individuals received reward for participation. The study was approved by Caltech Institutional Review Board.

**Evaluation of model and human performance**

Results from the testing data - for each of the model and human predictions - were processed as follows: In classification analysis, classified frames from the model/prediction were automatically sorted into one of four categories visible in the confusion matrix (polarized or non-polarized annotated true class, versus polarized or non-polarized predicted class). Cases in which the true class matched the predicted class were scored as an accurate prediction, and cases where the two classes did not match were scored as an inaccurate prediction. Population proportions of accurate results represent the proportion of accurate frames in the total population of frames. For time-smoothened data, the frames were first returned to time-order, after which the polarity onset point was determined by finding the point at which the prediction switched from an unpolarized majority to a polarized majority (as seen in Results). All frames after this polarity onset point were then classified as polarized, and all frames before this point were classified as unpolarized, therefore 'smoothening' out any anomalous predictions using time point information. For time point analysis, the polarity onset point (as determined from the smoothening process) was used. For each testing embryo time-lapse recording, the time discrepancy for the model/volunteer was calculated as the actual time difference (to the nearest second) between the predicted polarity onset frame and the annotated polarity onset frame, using the knowledge of the frame-to-frame time difference for each recording. Where no predicted onset frame was given within the allocated recording, for this analysis the frame immediately after the final frame of the time-lapse recording was used as the predicted onset of polarization. These time discrepancies for each embryo were used in pairwise comparisons.

**CAM attention map generation**

To identify focus areas of our ensemble model, we generated attention heat maps using the class activation mapping technique. To be specific, we multiplied each feature map passing through the global average pooling (GAP) layer of ResNet backbone with their corresponding weight connecting the GAP layer and the fully-connected layer. Then we added the weighted feature maps in an element-wise manner. Each weight tells us how much importance needs to be given to individual feature maps. The final weighted sum gives us a heat map of a particular class (in our case, the before/after polarity onset class), which indicates what pixels our model favors or dislikes to make the final prediction. The heat map size is the same as the one of feature maps. Therefore, to impose it on the input AIF DIC image, we scaled it to the size of the input image and finally got results shown in Fig. 3.

**Statistical analysis**

Image classification results were compared using a two-tailed *z*-test of two population proportions with significance classified for *p*-values as: $^*p < 0.05$, $^{**}p < 0.01$, $^{***}p < 0.001$, $^{****}p < 0.0001$ and not significant (NS). Time prediction discrepancies were compared using two-sided Wilcoxon matched-pairs signed-rank test since our testing data size is small and not guaranteed as normal. Significance was given for *p*-values as the same with the above. Further details are given with each result. Statistical analyses were performed using the statistics module in SciPy package with Python (https://docs.scipy.org/doc/scipy/reference/tutorial/stats.html). All the 95% confidence intervals were estimated by bootstrapping the testing dataset with 1000 replicates.

**Data Availability**

The testing dataset is available on https://github.com/Scott-Sheen/AI4Embryo for model validation use and academic purposes only. All other datasets generated and analyzed in the current study (including larger training image dataset) are available from the corresponding author (M.Z.-G) on reasonable request.

**Code Availability**

The training code for the single DCNNs and the testing code for the ensemble DL model are available at: https://github.com/Scott-Sheen/AI4Embryo.

**Acknowledgements**

We thank all colleagues in the C.Y. and M.Z.-G. labs for helpful suggestions and feedback. We also thank all human volunteers. We thank all funding sources: Wellcome Trust (098287/Z/12/Z) (MZG), Leverhulme Trust (RPG- 2018-085) (MZG), Open Philanthropy/Silicon Valley (MZG), Weston Havens Foundations (MZG), NIH R01 HD100456-01A1 (MZG), Rosen Bioengineering Center Pilot Research Grant Award (9900050) (CY, MZG), Medical Research Council (AL), Cambridge Vice Chancellor's Award Fund (AL).

**Author Contributions**

C.S. and A.L. were responsible for planning project directions, interpretation of results and optimization of the model. C.S. was responsible for the design of the model. M.Z. and A.L. were responsible for embryo recordings and assembly of the dataset. A.L. was responsible for annotating embryo images. The project was conceived by M.Z., M.Z.-G and C.Y. and supervised by M.Z.-G and C.Y. The manuscript was written by A.L., C.S., C.Y. and M.Z.-G. R.Z. edited the manuscript.

**Ethics Declarations**

*Competing Interests*

The authors declare that there are no competing interests.

*Institutional Review Board Statement*

All mouse experimental data was obtained in accordance with the Animals (Scientific Procedures) Act 1986 Amendment Regulations 2012, under project license by the University of Cambridge Animal Welfare and Ethical Review Body (AWERB). Reporting of animal experiments follows ARRIVE guidelines. Embryos were collected at the 2-cell stage from F1 females (C57BI6xCBA) mated with F1 studs, following super ovulation of the female: injection of 7.5 IU of pregnant mares' serum gonadotropin (PMSG; Intervet), followed by injection of 7.5 IU of human chorionic gonadotropin (HCG; Intervet) after 48 hours and immediately prior to mating.

All participants in the human trial provided informed consent before taking part in our study. They consented to allow their data to be used in the final analysis and all individuals received reward for participation. The human trial was approved by Caltech Institutional Review Board.

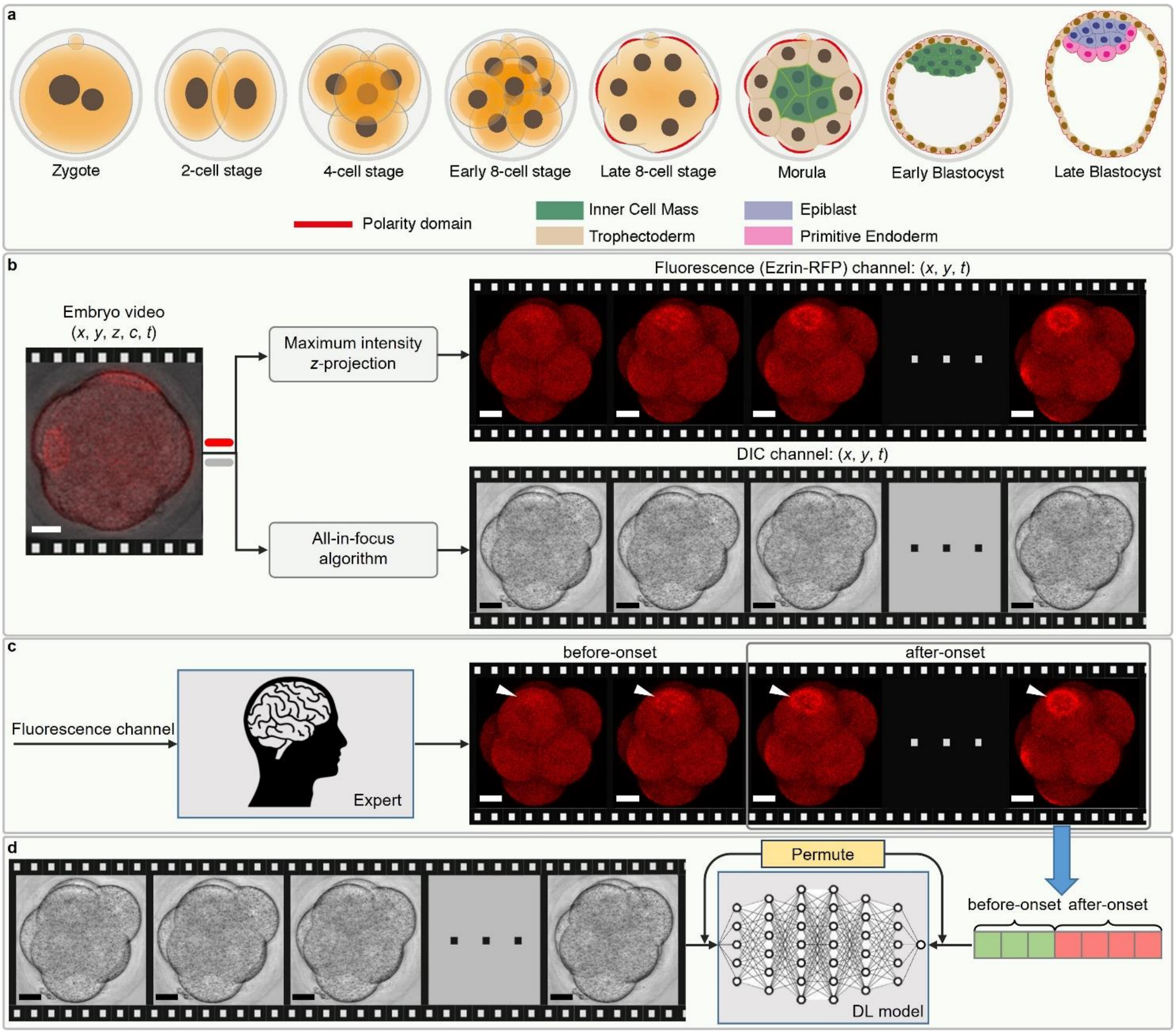

**Figure 1. Method to track and annotate polarity. a** Overview of mouse pre-implantation development, from the zygote stage at embryonic day 0 to the late blastocyst stage at embryonic day 4.5. At the late 8-cell stage, polarization takes place, as each blastomere gains a defined apical-basal axis of polarity indicated by the presence of an apical domain (red). **b** Data preprocessing of dual-channel 3D mouse embryo videos, each of which is a 5D tensor with the dimension of $x$, $y$, $z$, $c$ (channel), and $t$ (time). First, each video was split into a fluorescence (Ezrin-RFP) and DIC channel, visualized in red and gray respectively. Then, each channel was compressed along the $z$ dimension by different algorithms. The maximum intensity $z$-projection algorithm was applied for the fluorescence channel and DTCWT based AIF algorithm for the DIC channel to get the frame sequences. **c** Expert annotation on fluorescence frame sequences, where the time point of polarity onset is pinpointed. In the time sequence, the onset of polarization was defined as the frame in which the blastomere had a clear polarity ring or cap (closed) which took up at least 1/3 of the visible surface, or 1/3 of the cell surface curve if displayed side-on. Frames before this point were defined as before-onset, whilst frames including and after this point are defined as after-onset. **d** Supervised learning of a single DCNN model. The DIC frame sequences paired with the class labels from fluorescence annotation were permuted and used as the input and target of the supervised learning. Transfer learning from pre-trained weights on ImageNet database and data augmentation are utilized in the training of all DCNN models. Scale bar = 30 $\mu$m.

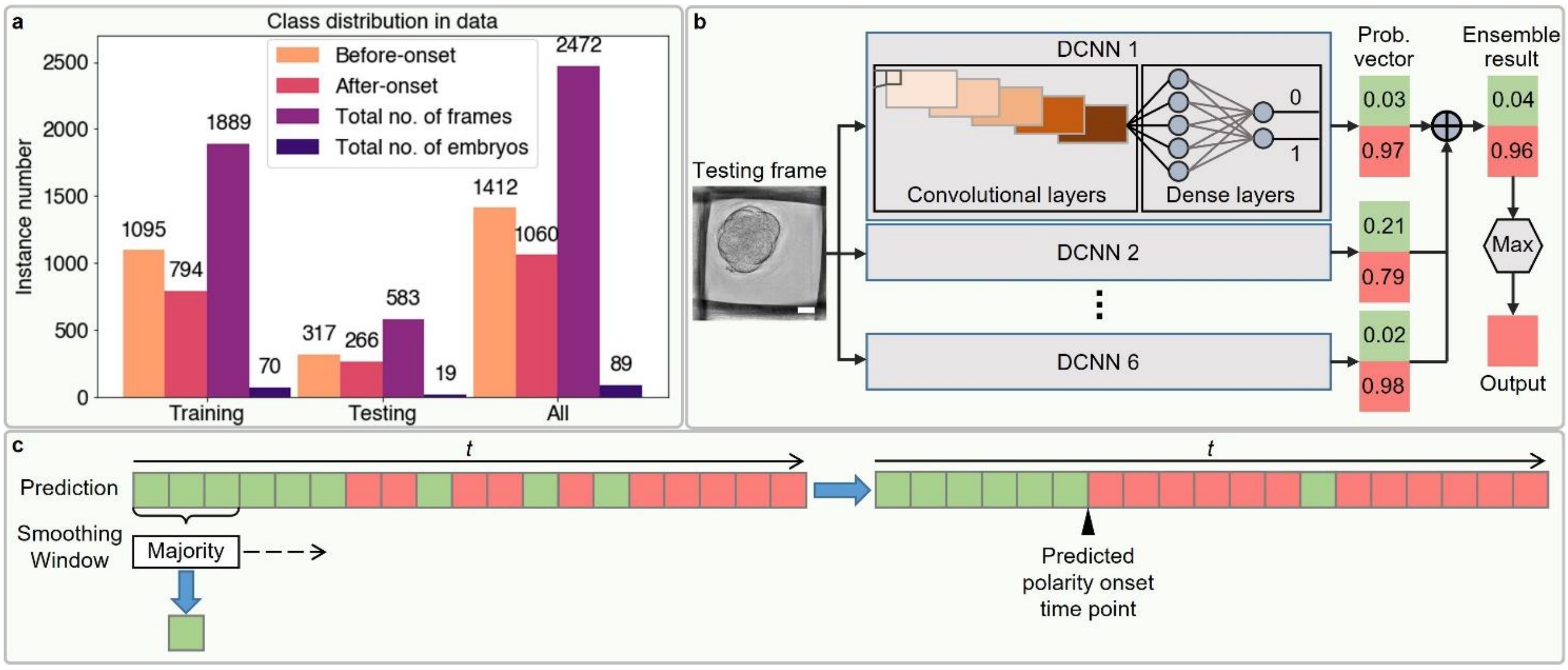

**Figure 2. An ensemble deep learning approach to predict embryo polarization from DIC images. a** Class distribution in the training/testing/whole dataset. **b** Ensemble learning on six DCNN models. The predicted probability vectors for two classes on a single testing frame by six DCNN models were averaged element-wisely and the class corresponding to the larger probability was used as the final predicted label. **c** Temporal smoothing on the predicted labels for each testing embryo's DIC frame sequence. The majority voting based smoothing window slid over the chronologically ordered binary labels. The window length is 3 and we kept the label at both ends untouched. Finally, the time index of first after-onset prediction was taken as the final prediction of polarity onset time point. Scale bar = 20 $\mu$m.

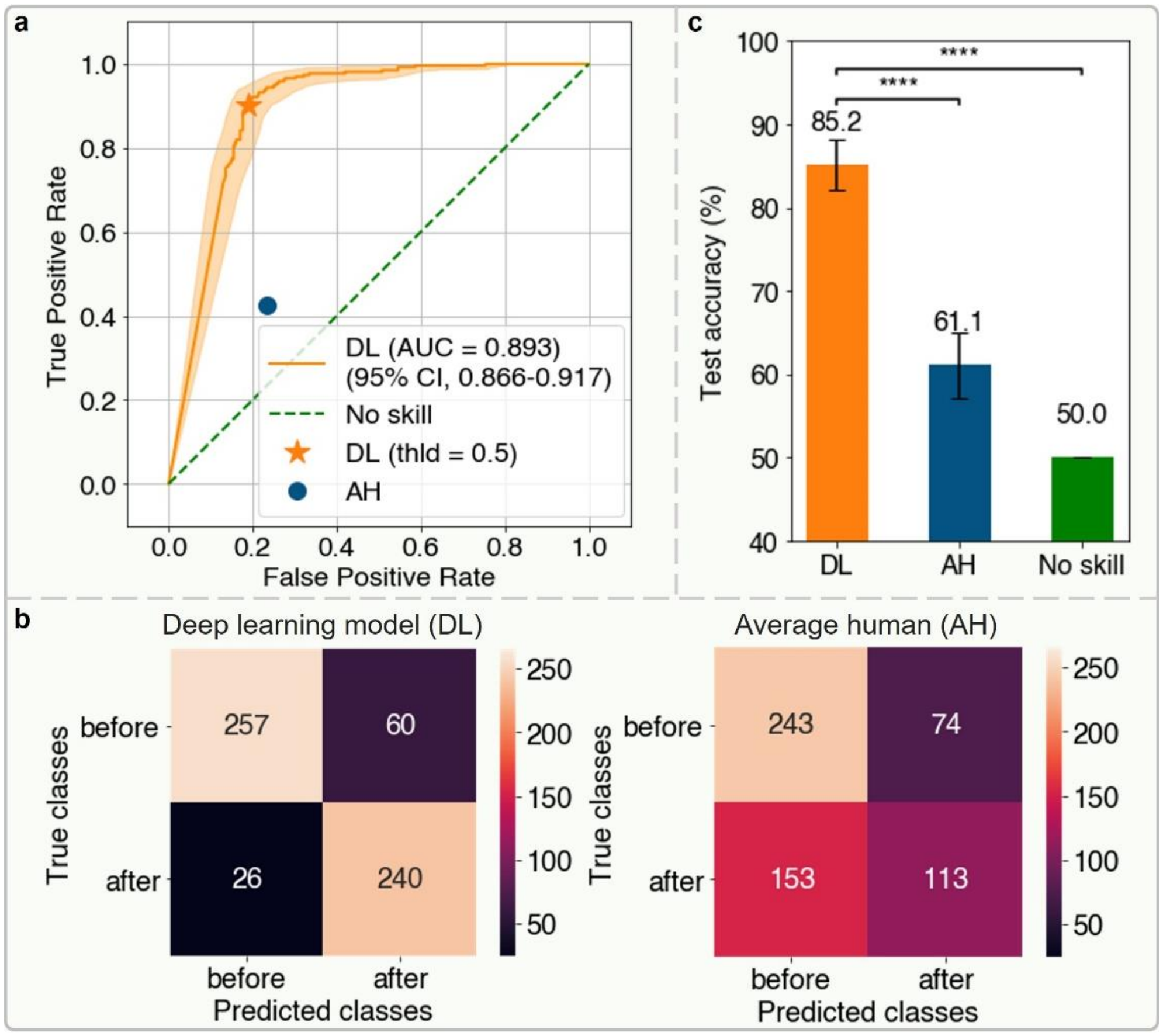

**Figure 3. Results of image classification task by the ensemble deep learning model and the average human. a** The receiver operating characteristic (ROC) curve of the performance of the ensemble deep learning (DL) model on testing frames. The 95% confidence intervals (CIs) of the ROC curve are indicated by the orange shaded area. The orange solid star represents the performance of the ensemble DL model with the default probability threshold of 0.5 to binarize its output and the dark blue solid circle represents the performance of the average human (AH), which is an aggregate result of six human volunteers' prediction. We applied majority voting to the six predictions on each testing frame to obtain the average human performance. If each prediction received three votes, we randomly assigned a prediction of before or after onset. **b** Confusion matrix of image classification on testing frames by the ensemble DL model with the binarization threshold of 0.5 and the average human. **c** Testing accuracy bar chart of the ensemble DL model and the average human compared with no skill (random predictions), where the error bars represent the 95% CI. The ensemble DL model significantly outperforms the average human, and the no skill predictions. $*p < 0.05$, $**p < 0.01$, $***p < 0.001$, $****p < 0.0001$, NS, not significant, two-sided $z$-test. All the 95% CIs are estimated by bootstrapping the testing dataset with 1000 replicates.

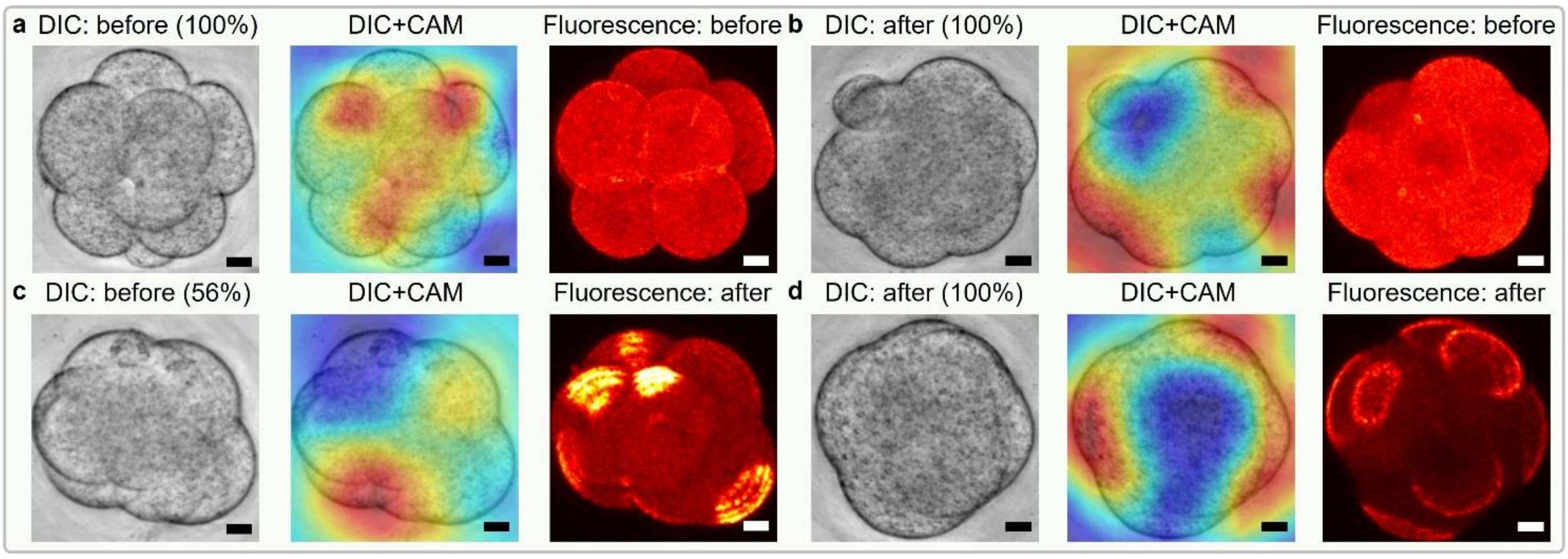

**Figure 4. Visualization of the decision-making by the ensemble deep learning model.** Heat maps obtained by the class activation mapping (CAM) technique highlight how the ensemble deep learning model attends the discriminative regions in the testing frame when giving the predicted class label. The red regions indicate positive focus of the model (in alignment with the predicted label) and the blue regions negative focus (in opposition to the predicted label). **a-d** correspond to four cases in confusion matrix, true negatives (TN), false positives (FP), false negative (FN), and true positives (TP), respectively. In each subfigure, from left to right are the testing DIC image, its overlay with the focus heat map, and its corresponding fluorescence channel image. On top of the test DIC image is the predicted label of the ensemble DL model with its confidence (from 0 to 100%). On top of the fluorescence image is the annotated label by the expert. All the heat maps show that our DL model either attends to the individual blastomeres or the inter-blastomere angles. For example, TP heat map **d** focuses on the truly polarized blastomeres. At a certain time-point, some blastomeres have started polarization but the others have not, as shown in the FN case **c**. This issue resulted in the DL model making a Type II error with low confidence in the case given. Scale bar = 20 $\mu$m.

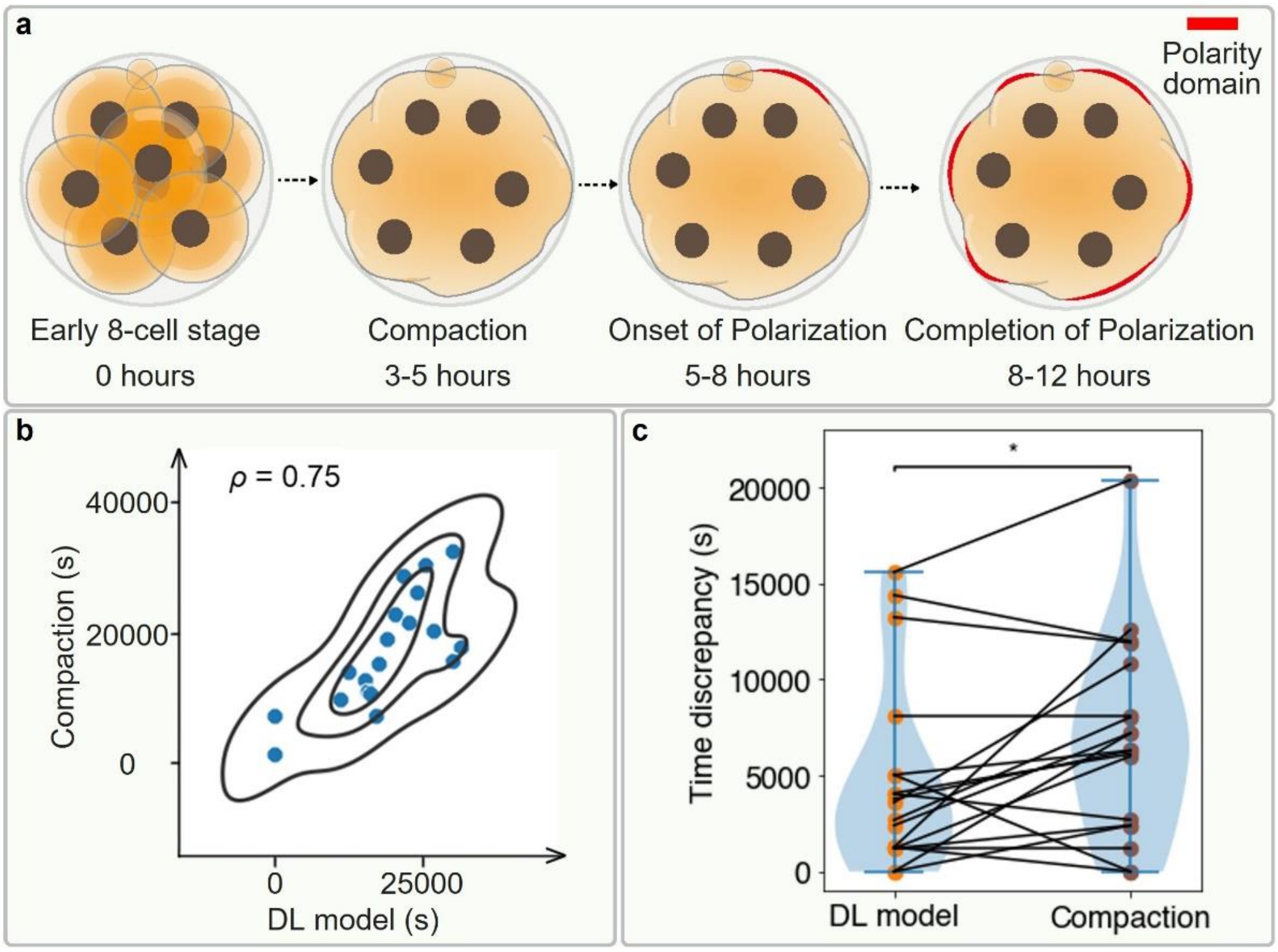

**Figure 5. Comparative analysis of the ensemble deep learning model prediction and the compaction-based prediction for polarization. a** Chronological order of compaction and polarization events during the 8-cell stage for a normal mouse embryo. **b** Correlation analysis between time points of DL model polarity prediction and compaction. The *x* and *y* coordinate are the predicted polarity onset time index of testing embryos (marked in blue solid balls) by the ensemble DL model and the annotated compaction time index, respectively. Their pairwise relationship shows a Pearson correlation coefficient ($\rho$) of 0.75. **c** Violin plot to visualize the time discrepancy between the annotated and the predicted polarity onset time index on 19 testing embryos by ensemble DL model and compaction proxy, overlaid with a slopegraph showing each testing embryo prediction time discrepancy in pair. From the kernel density estimate (blue shade) of violin plot and the connection line trends of slopegraph, we can tell that the prediction time discrepancy of DL model is significantly lower than the one of compaction proxy. The *p*-value is specified in the figure for $*p < 0.05$, $**p < 0.01$, $***p < 0.001$, $****p < 0.0001$, NS, not significant, two-sided Wilcoxon matched-pairs signed-rank test.

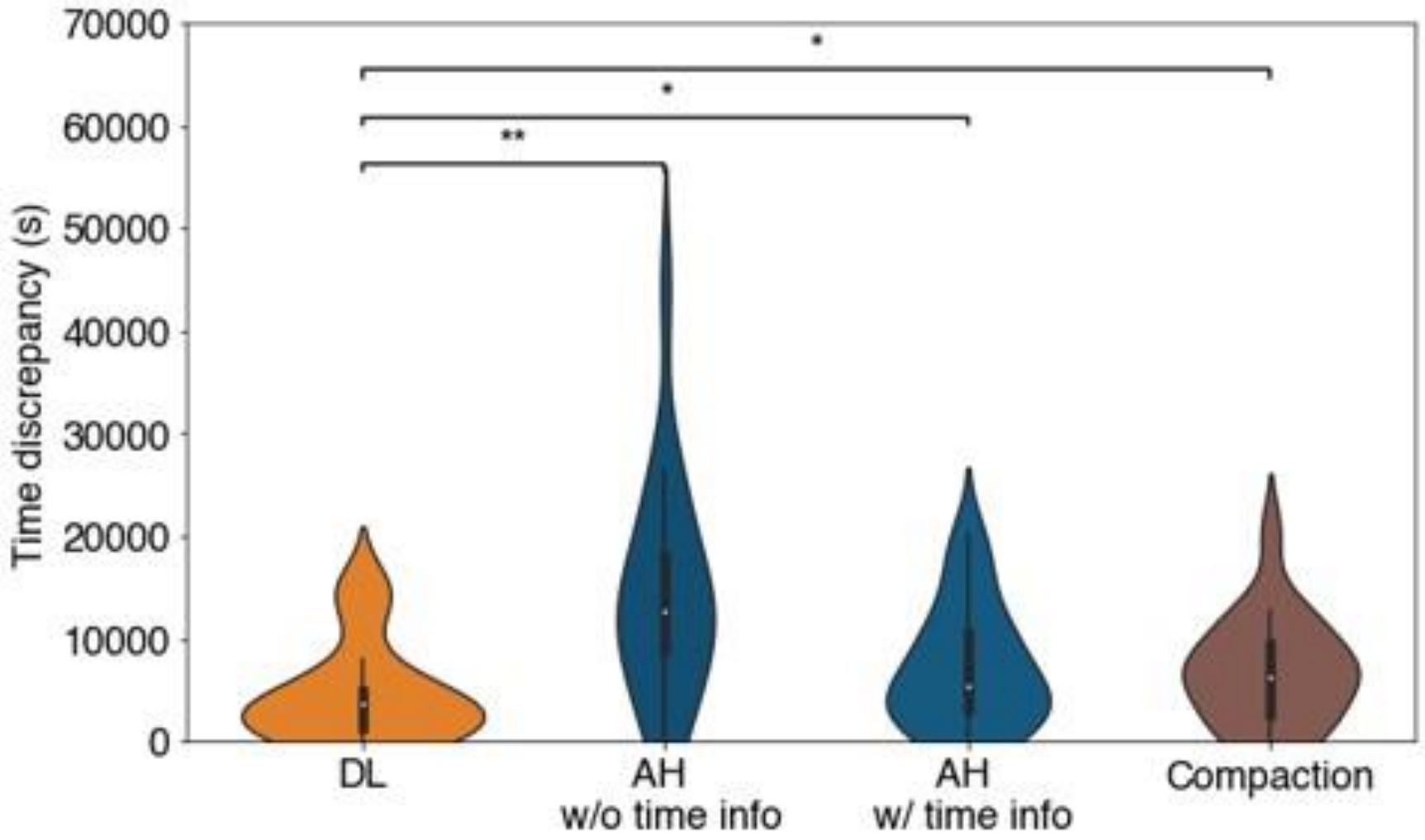

**Figure 6. Comparative analysis of the polarity onset time point prediction by the ensemble deep learning model, the average human and the compaction proxy.** Violin plot of time discrepancy between the annotated and the predicted polarity onset time index of 19 testing embryos by ensemble DL model, average human (AH) without/with time information and compaction proxy. AH without (w/o) time information (info) means that six human volunteers were given the randomized testing frames without any time information. Their predicted labels were then chronologically ordered for each testing embryo and temporally smoothened in the same manner as the ensemble DL model predictions. The mean discrepancy was taken from the six volunteers. AH with (w/) time information indicates that six human volunteers were given the chronologically ordered frames for each testing embryo. They directly estimated the polarity onset time point from these time sequences. Statistical analysis uses the ensemble DL model result as the reference to test their difference significance and the *p*-values are specified in the figure for $*p < 0.05$, $**p < 0.01$, $***p < 0.001$, $****p < 0.0001$, NS, not significant, two-sided Wilcoxon matched-pairs signed-rank test.

# Supporting Information

## Stain-free Detection of Embryo Polarization using Deep Learning

Cheng Shen[1,+], Adiyant Lamba[2,+], Meng Zhu[2,3], Ray Zhang[4], Changhuei Yang[1,5,*], and Magdalena Zernicka Goetz[2,5,*]

[1] Department of Electrical Engineering, California Institute of Technology, Pasadena, CA, USA

[2] Mammalian Embryo and Stem Cell Group, Department of Physiology, Development and Neuroscience, University of Cambridge, Downing Street, Cambridge, CB2 3EG, UK

[3] Blavatnik Institute, Harvard Medical School, Department of Genetics, Boston, MA 02115, USA

[4] Department of Pathology and Immunology, Washington University School of Medicine, St. Louis, MO, USA

[5] Division of Biology and Biological Engineering, California Institute of Technology, Pasadena, CA, USA

[+] These authors contributed equally

[*] co-corresponding authors: chyang@caltech.edu, magdaz@caltech.edu

**Including:**

Figures S1 to S6

Table S1

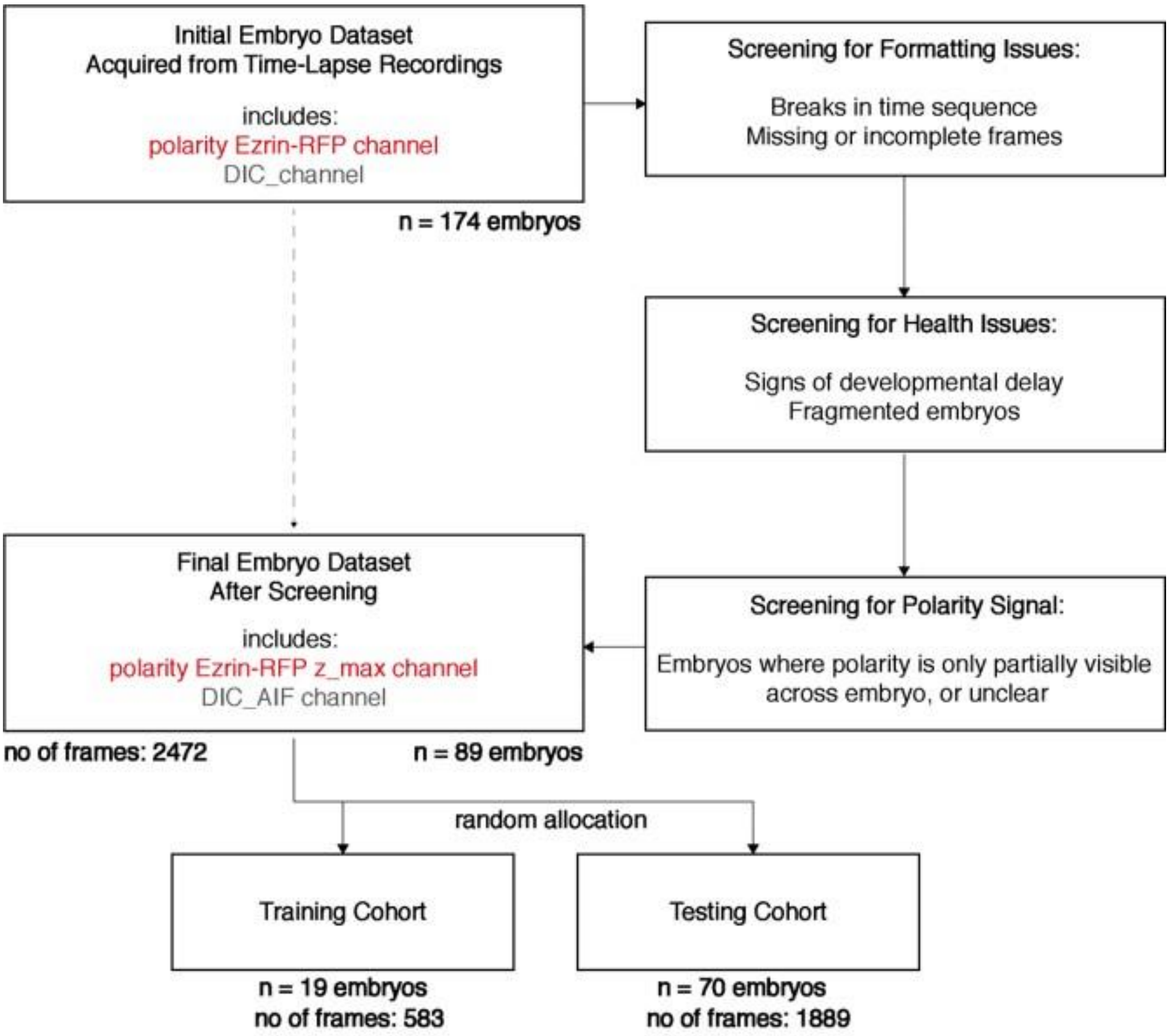

**Figure S1. Flowchart indicating initial cleaning and pre-processing of data.**
We analyzed 174 mouse embryo time-lapse recordings from dual-modal confocal microscope imaging, containing a DIC channel and a fluorescent polarity-indicating channel. After screening for image and embryo development quality, 89 embryos were left, with each channel compressed along the *z* axis using different algorithms and then randomly split into a training and testing cohort.

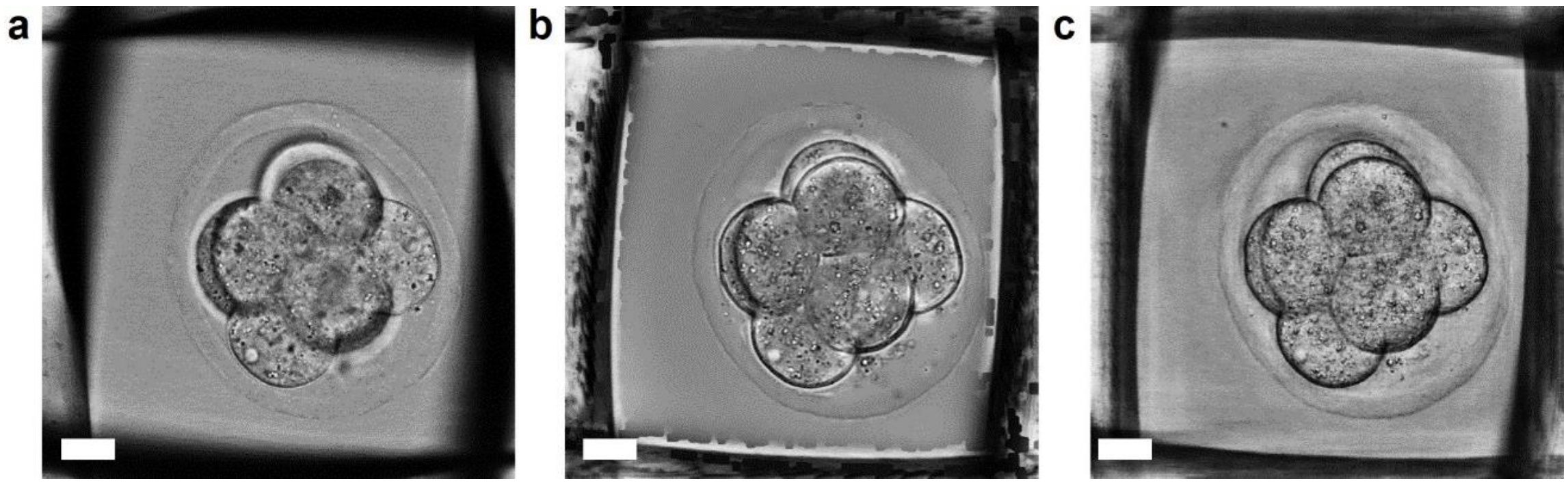

**Figure S2. Comparison among different input image formats.**
Previous deep learning studies on embryo development used a single *z* slice image, in most cases the middle plane - see (a). However, this resulted in some blastomeres being highly defocused and blurred. The traditional all-in-focus algorithm based on variance metric (b) can bring all the blastomeres into focus in a single 2D image but also result in some artifacts. Thus, we proposed to utilize the all-in-focus algorithm based on dual tree complex wavelet transform (c). Scale bar = 20 $\mu$m.

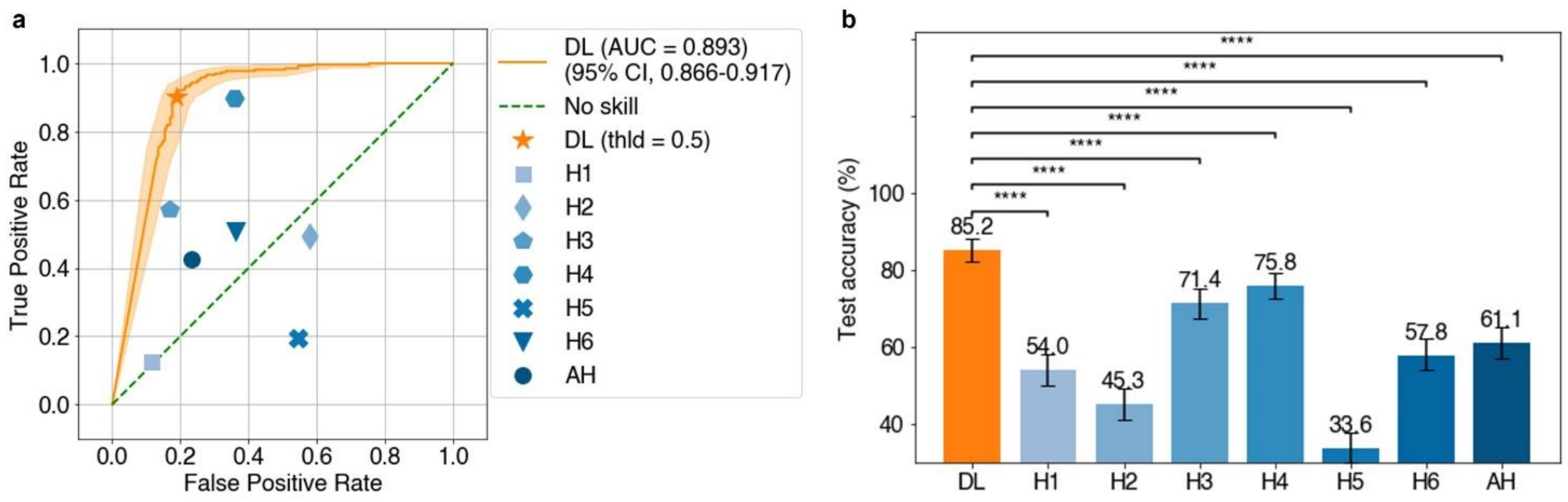

**Figure S3. Results of image classification task by the ensemble deep learning model, six human volunteers, and their average.**

**a** The receiver operating characteristic (ROC) curve of the performance of the ensemble DL model on testing frames. The 95% confidence intervals (CIs) of the ROC curve were indicated as the orange shading area. The orange solid star represents the performance of the ensemble DL model with the default probability threshold (thld) of 0.5 to binarize its output and the blue markers with different shapes and saturation represent the performance of six human volunteers and their average (AH). We applied majority voting to the six predictions on each testing frame to obtain an average human performance. In the case of a tie, we randomly assigned the prediction of before or after onset. **b** Testing accuracy bar chart of the ensemble DL model, six human volunteers and their average (AH), where the error bars represent the 95% CI. The ensemble DL model significantly outperforms each individual human and the average human. *$p < 0.05$, **$p < 0.01$, ***$p < 0.001$, ****$p < 0.0001$, NS, not significant, two-sided $z$-test used in all cases. All the 95% CIs are estimated by bootstrapping the testing dataset with 1000 replicates.

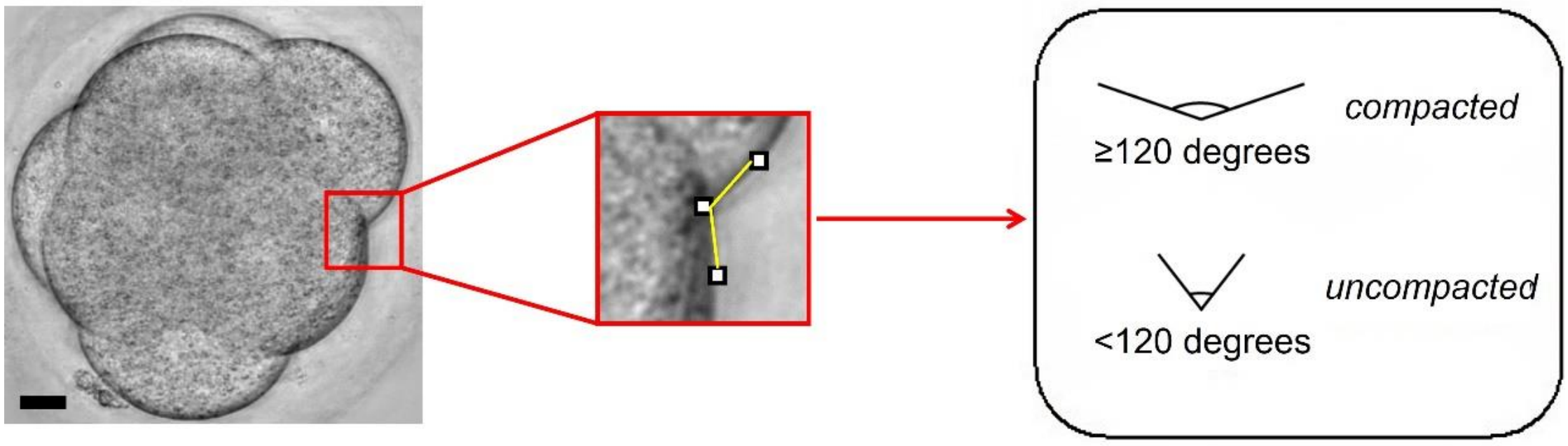

**Figure S4. Criteria for annotating compaction.**
We measured compaction using the inter-blastomere angle. The first time point at which the minimal inter-blastomere angle was ≥ 120 degrees was defined as the compaction point. All frames including and succeeding this point were defined as compacted, whilst all frames prior to this point were defined as uncompacted. Scale bar = 10 $\mu$m.

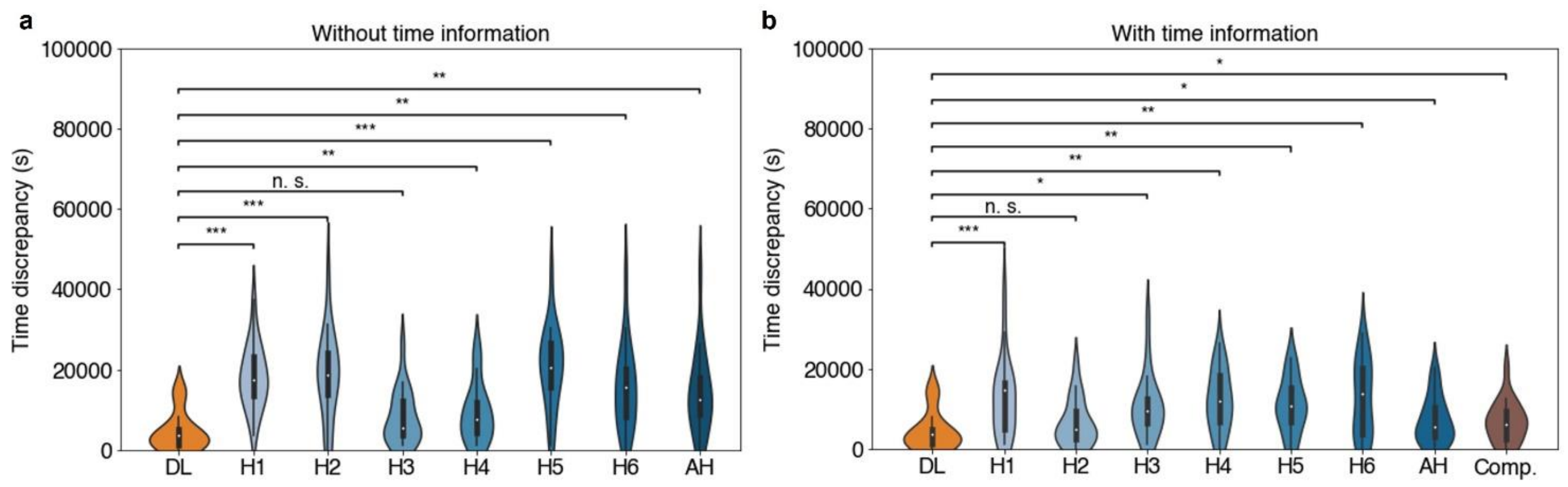

**Figure S5. Comparative analysis of the polarity onset time point prediction by the ensemble deep learning model, six human volunteers, their average, and the compaction proxy.**

The violin plots represent time discrepancies between the annotated and the predicted polarity onset time index of the 19 testing embryos by the ensemble DL model, each of six human volunteers (H1-H6), their average (AH), and the compaction proxy (Comp.). **a** Six human volunteers were given the randomized testing frames without any time information. Their predicted labels were then chronologically ordered for each testing embryo and temporally smoothened to extract the polarity onset time point prediction, as shown in Fig. 1g. Their average result was processed in the same way. **b** Six humans were given the chronologically ordered frames for each testing embryo. They directly estimated the polarity onset time point. Their average result was the arithmetic mean of predicted time indexes for each testing embryo. Comparison between the ensemble DL model and each human is given in the figure. *$p < 0.05$, **$p < 0.01$, ***$p < 0.001$, ****$p < 0.0001$, NS, not significant, two-sided Wilcoxon matched-pairs signed-rank test.

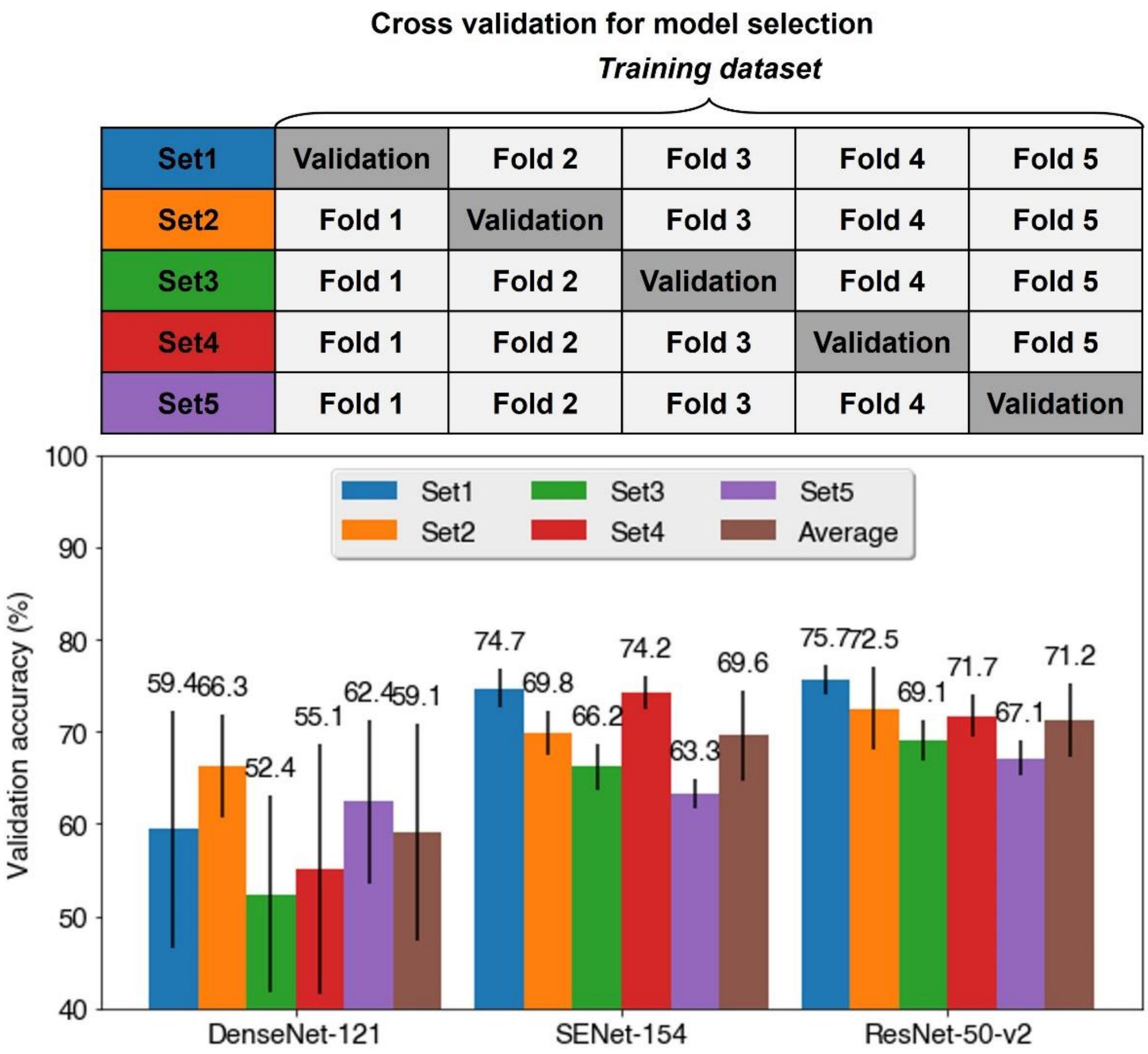

**Figure S6. Comparison among the performance of different image classification DCNN backbones.** Five-fold cross validation was adopted here to make model selection. The training dataset was evenly split into 5 folds. Then, three backbone models learnt on four folds and were validated on the remaining one. To minimize the variance brought by the optimization setting, we repeated each numerical experiment 5 times. We summarized the validation accuracy of three models on each fold and the total average validation accuracy in the bar chart, where the error bars represent their standard deviation. Both validation accuracy and their standard deviation shows that ResNet-50-v2 is the optimal backbone for our task.

| Metrics | DL | H1 | H2 | H3 | H4 | H5 | H6 | AH |
|---|---|---|---|---|---|---|---|---|
| Sensitivity (%) | **90.2** (86.1, 93.8) | **12.8** (8.8, 16.8) | **49.2** (43.3, 55.0) | **57.1** (51.0, 63.1) | **89.8** (85.9, 93.4) | **19.5** (14.6, 24.3) | **50.8** (44.5, 56.7) | **42.5** (36.4, 48.5) |
| Specificity (%) | **81.1** (76.2, 85.4) | **88.6** (85.0, 92.2) | **42.0** (36.8, 47.3) | **83.3** (79.0, 87.2) | **64.0** (59.0, 69.3) | **45.4** (40.3, 50.8) | **63.7** (58.4, 69.2) | **76.7** (71.9, 81.1) |
| PPV (%) | **80.0** (75.1, 84.5) | **48.6** (36.2, 60.9) | **41.6** (35.7, 46.9) | **74.1** (68.1, 79.9) | **67.7** (63.0, 72.6) | **23.1** (17.7, 28.9) | **54.0** (48.2, 60.1) | **60.4** (53.6, 67.2) |
| NPV (%) | **90.8** (87.2, 94.2) | **54.8** (50.4, 58.9) | **49.6** (43.7, 55.6) | **69.8** (65.4, 74.5) | **88.3** (83.8, 92.5) | **40.2** (35.1, 45.4) | **60.7** (55.4, 66.1) | **61.4** (56.3, 66.2) |
| Accuracy (%) | **85.2** (82.2, 88.2) | **54.0** (49.9, 58.0) | **45.3** (41.0, 49.2) | **71.4** (67.4, 75.0) | **75.8** (72.4, 79.2) | **33.6** (30.0, 37.6) | **57.8** (54.0, 62.1) | **61.1** (57.1, 65.0) |

**Table S1. The performance of the ensemble deep learning (DL) model, six human volunteers (H1-H6) and their average (AH) on the testing dataset.**

95% confidence intervals are included in brackets. They are estimated by bootstrapping the testing dataset with 1000 replicates. The highest value of each metric achieved by all predictors is marked in bold red. PPV, positive predictive value; NPV negative predictive value. H1-H3 are females and H4-H6 are males.